\documentclass[conference]{IEEEtran}
\usepackage{graphicx}
\usepackage{etex}
\usepackage{booktabs}
\usepackage[utf8]{inputenc}
\usepackage[ruled,vlined]{algorithm2e}
\usepackage[T1]{fontenc}
\usepackage{microtype}
\usepackage{graphicx}
\usepackage{paralist}
\usepackage{tabularx}
\usepackage{balance}
\usepackage{multicol}
\usepackage{multirow}
\usepackage{pbox}
\usepackage{enumitem}	
\usepackage{amssymb}
\usepackage{colortbl}
\usepackage{pifont}
\usepackage{xspace}
\usepackage{url}
\usepackage{tikz}
\usepackage[TABBOTCAP]{subfigure}
\usepackage{amsmath}
\usepackage{threeparttable}
\usepackage{pbox}
\usepackage{ragged2e}
\usepackage{stfloats}
\usepackage{flushend}

\ifCLASSINFOpdf
\else
\fi
\begin{document}
%
\title{Automatic Traceability Maintenance via Machine Learning Classification}

\author{Anonymous Authors}

%
\author{\IEEEauthorblockN{Chris Mills,
Javier Escobar-Avila,
Sonia Haiduc
}
\IEEEauthorblockA{Department of Computer Science - Florida State University\\Tallahassee, FL, USA\\ \{cmills, escobara, shaiduc\}@cs.fsu.edu}
}


\newcommand\approach{TRAIL}

\maketitle
\begin{abstract}
Previous studies have shown that \textit{software traceability}, the ability to link together related artifacts from different sources within a project (e.g., source code, use cases, documentation, etc.), improves project outcomes by assisting developers and other stakeholders with common tasks such as impact analysis, concept location, etc. 
Establishing traceability links in a software system is an important and costly task, but only half the struggle. As the project undergoes maintenance and evolution, new artifacts are added and existing ones are changed, resulting in outdated traceability information. Therefore, specific steps need to be taken to make sure that traceability links are maintained in tandem with the rest of the project.

In this paper we address this problem and propose a novel approach called \approach{} for maintaining traceability information in a system. The novelty of \approach{} stands in the fact that it leverages previously captured knowledge about project traceability to train a machine learning classifier which can then be used to derive new traceability links and update existing ones. We evaluated \approach{} on 11 commonly used traceability datasets from six software systems and compared it to seven popular Information Retrieval (IR) techniques including the most common approaches used in previous work. The results indicate that \approach{} outperforms all IR approaches in terms of 
precision, recall, and F-score.


\end{abstract}


%
\IEEEpeerreviewmaketitle

\section{Introduction}
Software systems are comprised of information stored in various artifacts such as source code, bug reports, requirements specifications, use and test cases, interaction diagrams, and user documentation among others. Traceability Link Recovery (TLR) is the software engineering task focused on establishing bidirectional links (i.e., \textit{traceability links}) between related  artifacts of different types in order to provide developers and stakeholders a detailed picture of how a system is constructed and the relationships between system components. The resulting software traceability naturally supports other tasks such as concept location, impact analysis, program comprehension, verifying test coverage, ensuring that system and regulatory requirements are met, etc. and has been proven to be useful in practice \cite{bouillon2013survey, mader2015developers, mader2016preventing}.

Establishing traceability links between the artifacts of a system is extremely arduous and error-prone when performed manually. This has led to a large body of research proposing techniques that aid developers with this task \cite{borg2014recovering, marcus2005recovery, abadi2008traceability, asuncion2010software, kong2011proximity}. However, establishing traceability links at one point in the lifetime of a software system is only part of the struggle. As the system evolves over time, some artifacts get deleted, new artifacts are added that have not yet been linked to others, and substantial changes 
can break existing links. Therefore, the benefits of software traceability can quickly degrade unless this information is updated and maintained in tandem with the evolving software artifacts \cite{james1997automatic, weidenhaupt1998scenarios, antoniol2000traceability}.

In this paper we address this problem and propose a novel approach for traceability maintenance. Our approach, called \approach{} (TRAceability lInk cLassifier), uses historically collected traceability information (i.e., existing traceability links between pairs of artifacts) to train a machine learning classifier which is then able to classify the link between any new or existing pair of artifacts as \textit{valid} (i.e., the two artifacts are related) or \textit{invalid} (i.e., the two artifacts are unrelated). Since \approach{} relies only on the features of the two artifacts in order to determine the validity of the link between them, it is able to classify links between brand new artifacts introduced in the system, as well as to reassess existing links, when the artifacts involved have changed. To the best of our knowledge \approach{} is the first approach that  requires neither human intervention nor a set of predefined rules for these tasks.

We evaluate \approach{} on a set of 11 datasets from six software systems which are commonly used in traceability studies \cite{hayes2003improving, DBLP:journals/tse/HayesDS06, cleland2010machine, ali2011requirements, gethers2011integrating}. We also compare \approach{} to Information Retrieval (IR) techniques, which are the most popular type of technique used in traceability link recovery and maintenance \cite{borg2014recovering}. In particular, we compare \approach{} to seven IR approaches, including the most common approaches used in previous work \cite{marcus2005recovery, abadi2008traceability, asuncion2010software, oliveto2010equivalence}. The results of our evaluation reveal that \approach{} is able to achieve high precision, recall and F-measure, significantly outperforming all IR approaches. 


The main contributions of this paper are:
\begin{enumerate}
	\item A novel approach to traceability maintenance called \approach{} which uses historical trace information to train a machine learning classification algorithm that predicts if a new or updated traceability link is valid or invalid.
	\item An empirical derivation of \approach{} that shows which feature selection, balancing technique, and classification algorithm provide the best performance.
	\item An empirical evaluation of \approach{} on 11 popular traceability datasets, which indicates that \approach{} is able to achieve high precision, recall, and F-measure.
	\item A comparison of \approach{} with seven IR approaches previously used in traceability link recovery and maintenance, which shows that \approach{} significantly outperforms IR in terms of precision, recall, and F-measure. 
\end{enumerate}

The remainder of the paper is structured as follows: section \ref{sec:approach} presents our approach; sections \ref{sec:methodology} and \ref{sec:results} describe the evaluation we performed and a discussion of its results; section \ref{sec:threats} discusses threats to validity; section \ref{sec:related} provides an overview of related work; and section \ref{sec:conclusion} summarizes our conclusions and presents ideas for future research.

\section{Approach}
\label{sec:approach}
In this section we introduce our novel approach to traceability link maintenance, called \approach{} (TRAceability lInk cLassifier). Different from existing approaches, \approach{} leverages historical traceability information (i.e., pre-existing traceability links) to infer how artifacts should be linked. \approach{} transforms the traceability maintenance problem into a binary classification problem where the links between pairs of artifacts are classified as being valid or invalid. It employs machine learning algorithms that use features derived from the existing traceability links to infer statistical patterns that differentiate the valid and invalid links between artifacts. Due to its construction, \approach{} can assess both existing and new pairs artifacts in the system, which represents an advantage over previous approaches. 

Rather than setting in stone the algorithms, features, and configurations, we designed  \approach{} as a framework where individual components can be replaced by others that perform a similar task. This ensures that \approach{} is adaptable and configurable to specific needs that could arise in other domains or applications. In the following subsections we first present the framework itself at a high level, and then describe the specific implementation we considered in our study.

The \approach{} framework consists of four main components: 
\begin{enumerate}
\item A set of \textit{features} that are meant to represent the potential traceability links between software artifacts;
\item A \textit{feature selection} component that extracts representative features from all the ones initially considered, in order to reduce the dimensionality, avoid overfitting, and improve the learning and generalization of the classifier.
\item A \textit{data balancing} component that, if needed, rebalances the training data between the valid and invalid classes of links, to ensure better model training and performance;
\item A \textit{machine learning classification algorithm} that is used to determine which potential links are valid and which are invalid. 
\end{enumerate}

\begin{figure}
	\centering
	\includegraphics[width=1.0\linewidth, trim=0 0 15 2,clip]{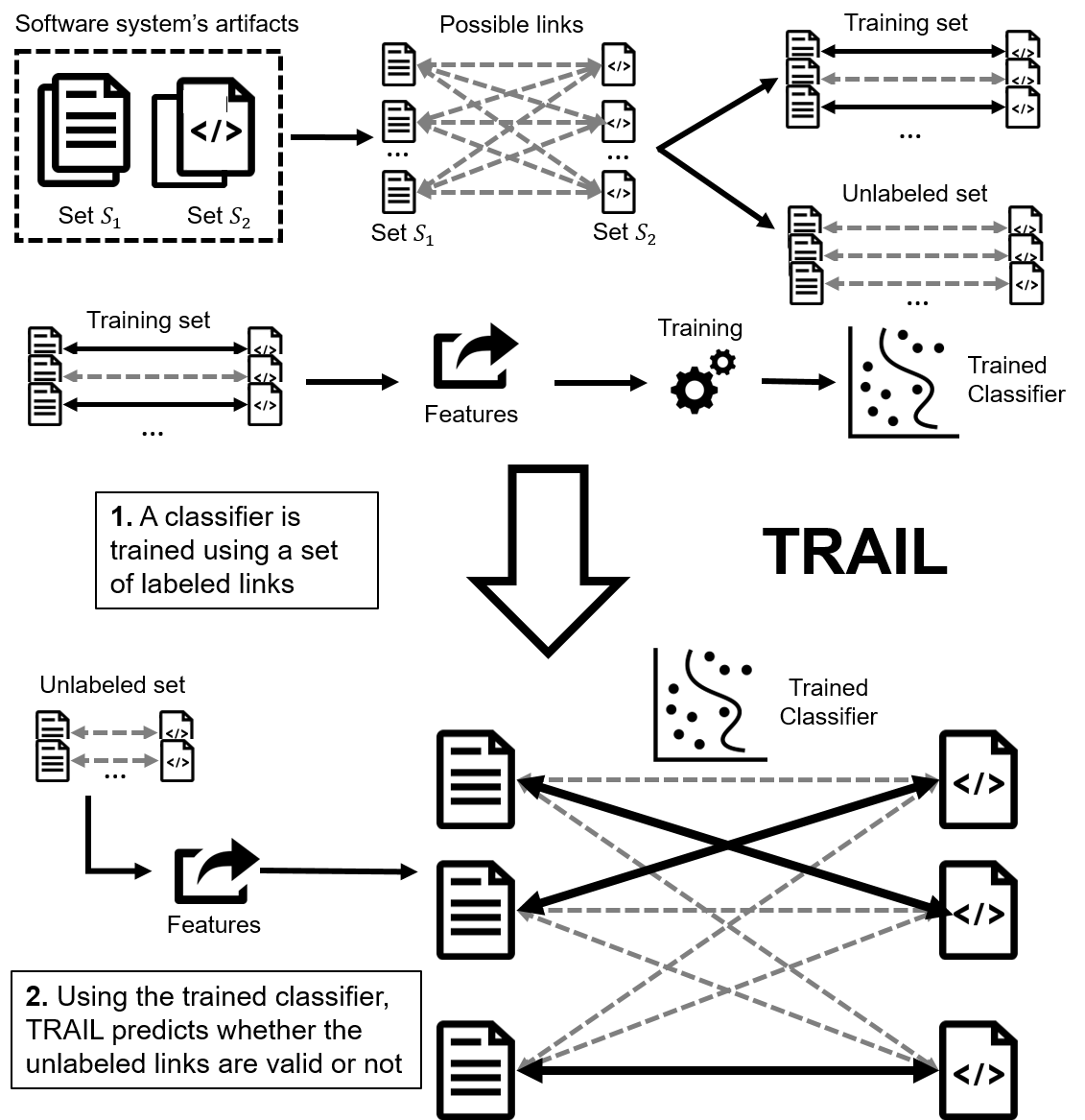}
	\caption{The general \approach{} framework for traceability link classification}
	\label{fig:approach}
	\vspace{-0.7cm}
\end{figure}

Based on the above components, a few steps need to be performed in order to instantiate and use \approach{}. First, feature engineering is employed to establish a set of features that can be used to represent potential traceability links (section \ref{subsec:features} describes our features). Then a feature selection algorithm is selected and applied to extract only the most relevant features (more details in section \ref{subsec:selection}). Next, because the approach represents all possible links between two sets of artifacts, the training data is expected to be highly imbalanced (i.e., given two sets of artifacts, there are usually much fewer \textit{valid} links than \textit{invalid} ones between them). To address this situation, a rebalancing technique is selected and applied to the data prior to training the classifier; this is further discussed in section \ref{subsec:imbalance}. Finally, a classification algorithm is chosen and trained with the resulting data and can then be used to classify new links. Figure \ref{fig:approach} shows an overview of how an instance of \approach{} is trained and used to predict the validity of traceability links.

Note that when implementing \approach{} the implementor is at liberty to design a variation that meets her specific needs by choosing a completely custom set of features, feature selection algorithm, rebalancing technique, and classifier. Moreover, \approach{} can easily be extended with emerging techniques in machine learning, which allows it to implicitly benefit from a rapidly evolving discipline and research area. In the sub-sections that follow, we introduce the specific settings we used to instantiate \approach{} for the study in this paper. 

\subsection{Features Representing the Links}
\label{subsec:features}
\approach{} considers all of the possible links that could exist between two given sets of software artifacts and predicts each to be either valid (i.e., the artifacts are related) or invalid (i.e., the artifacts are not related). Formally, if we have two artifact sets $S_{1}$ and $S_{2}$, the approach predicts the validity of each element in the Cartesian product $S_1 \times S_2$. That is, for each artifact $s\in S_1$, we predict the validity of the potential link that exists between $s$ and some $s'\in S_{2}$.  In order to perform this classification, we must have a set of features that define an internal, vector representation for our classifier. The features used in our implementations of \approach{} can be separated into three categories: IR-based, query quality (QQ), and document statistics features.
	
\subsubsection{IR-based Features}\label{sec:ir-features}
\label{IR_features}
While IR-based techniques do not provide a silver bullet for traceability, their ability to capture the semantic similarity between software artifacts has long been established. Therefore, while our approach does not use IR to propose lists of candidate links to a human, we still leverage the power of IR in the first set of features. 

Given two artifact sets $S_{1}$ and $S_{2}$, and a possible link between artifacts $s1 \in  S_{1}$ and $s2 \in S_{2}$, we capture the strength of this link based on IR using two metrics. First, we use $s1$ as a query and the artifacts in $S_{2}$ as the corpus. After running $s1$ as a query through an IR engine, we capture the rank at which $s2$ appears in the list of results as the first metric. Then we repeat the procedure, this time considering $s2$ as the query, $S_{1}$ as the corpus, and capturing the rank of $s1$ in the list of results as the second metric. Therefore, the representation of each potential link in \approach{} considers trace link recovery from either direction (i.e., using either artifact in the link as a query), whereas traditional IR approaches consider only a single direction. This is an important distinction, as previous work has shown that the choice of retrieval direction impacts retrieval performance specifically for traceability \cite{mills2017impact}.

In this study, we use seven different IR approaches to compute each of the two aforementioned metrics, resulting in a total of \textit{14 different IR-based features for each possible link}. The IR approaches we use are: Vector Space Model with TF-IDF weighting and cosine similarity, Okapi BM25, Jensen Shannon, Latent Semantic Analysis, Latent Dirichlet Allocation, and two approaches based on smoothing methods for language models: Dirichlet and Jelenik-Mercer \cite{zhai2001study}. Next we provide a short description of each of these approaches.

\emph{Vector Space Model (VSM):} In the VSM, software artifacts/documents are represented using a \textit{term-by-document} matrix. Each element of the matrix stores the importance of a term in the document and corpus, expressed using its \emph{term frequency-inverse document frequency} (TF-IDF). Since each document is represented as a vector, we can compute the similarity between documents using cosine similarity:

\vspace{-.4cm}
\begin{align}\label{eq:cosine}
sim_{cosine}(d_1,d_2)=\frac{\sum_{i=1}^{t}{d_{1,i}{d_{2,i}}}}{\sum_{i=1}^{t}{d_{1,i}^2} \cdot \sum_{i=1}^{t}{d_{2,i}^2}}
\end{align}

where $d_1$ and $d_2$ are two documents in the corpus, $t$ is the total number of unique terms in the corpus, $d_{1,i}$ and $d_{2,i}$ are the TF-IDF weights of the $ith-$ term in each document. 
	
\emph{Latent Semantic Analysis (LSA):} LSA \cite{deerwester1990lsi} is able to capture information about the co-occurrence of terms, addressing the synonymy and polysemy problems that VSM is not able to. LSA uses Singular Value Decomposition to decompose the term-by-document matrix into three matrices, one of them containing a matrix of singular values ($S_0$). By taking the $k$ largest values from $S_0$, a new term-by-document matrix can be reconstructed with low dimensionality and information about the association between terms. The document representation in the reconstructed matrix together with cosine similarity is used to compute similarity between documents.

\emph{Latent Dirichlet Allocation (LDA):} LDA is a generative probabilistic model that represents each document as a mixture of latent topics, and each topic as a distribution over words in the corpus \cite{blei2003lda}. After using LDA, each document is represented as a vector of probabilities, each one describing the probability of a topic to appear in the document. We use this representation, together with Hellinger distance, to compute the similarity between documents. Specifically, we define the similarity between two LDA document representations $d_1$ and $d_2$ as:
\vspace{-0.4cm}
\begin{align}
	sim_{LDA}(d_1,d_2) = 1 - \frac{1}{\sqrt{2}} \|(\sqrt{d_1} - \sqrt{d_2})\|
\end{align}
\vspace{-0.4cm}

Since our goal is to accurately estimate the similarity between two documents in the topic space, we configured the LDA implementation\footnote{We use the LDA implementation offered by the R package \emph{topicmodels}.} with a large number of topics (250). 

\emph{Jensen-Shannon (JS):} The JS model also represents software artifacts as probability distributions over terms in the corpus via hypothesis testing techniques \cite{abadi2008traceability}. Given the JS representation of two documents $d_1$ and $d_2$, their similarity is computed as follows:

\vspace{-0.5cm}
\begin{align}	
		sim_{JS}(d_1,d_2) & = 1 - \left [ H \left (  \frac{d_1+d_2}{2} \right ) - \frac{H(d_1)+H(d_2)}{2}  \right ] \nonumber \\
		H(d) & = \sum_{w \in W}{-P(w)\cdot \log P(w)} 	
\end{align}
\vspace{-0.4cm}

where $w$ is a word in document $d$, $W$ is the set of unique words in document $d$, and $P(w)$ is the probability of word $w$ appearing in document $d$.

\emph{Okapi BM25:} The BM25 model scores each document in the corpus based on the query terms appearing in it. The scoring function is the following \cite{manning2008ir}:
\begin{equation}
	\begin{split}
		score_{BM25}(q,d) = &\left ( \sum_{t \in q}{\log \left[ \frac{N}{df_t} \right]} \right ) \cdot \\ 
							&\frac{(\lambda+1)\cdot tf_{t,d}}{tf_{t,d} \cdot \lambda \left ( (1-b) +b \cdot \left ( \frac{L_d}{L_{ave}} \right )  \right ) }
	\end{split}
\end{equation} 

where $q$ is the query, $d$ is a document in the corpus, $N$ is the number of documents in the corpus, $df_t$ is the number of documents the term $t$ appears in, $tf_{t,d}$ is the term frequency of term $t$ in document $d$, $L_d$ is the length of document $d$ (in number of words), $L_{ave}$ is the average document length in the corpus, $b$ is a parameter used to control how much effect field-length normalization should have, and $\lambda$ is a positive parameter that calibrates the document term frequency scaling. We used Lucene's\footnote{\url{https://lucene.apache.org/} Version 6.4.1} default implementation of BM25, which utilizes $\lambda=1.2$ and $b=0.75$.

\emph{Language Model with Dirichlet (LM-Dirichlet) and Jelinek-Mercer smoothing (LM-JM):} These two models first define a language model for each document in the corpus, and then rank the documents according to the probability that the language model of each document $d$ generates a query $q$. Both models use \emph{smoothing} to improve accuracy by adjusting the maximum estimator of a language model\cite{zhai2001study}. 

A general form of a smoothed language model is the following: 

\begin{equation}
	p(q,d) = \prod_{w \in q}{p(q_i|d)}
\end{equation}

where

\begin{equation*}
	p(q_i|d) = 
	\begin{cases}
		p_{smooth}(q_i|d) \text{ if word $q_i$ is seen in $d$} \\
		\alpha \cdot p(q_i|C) \text{ otherwise}
	\end{cases}
\end{equation*}

$\alpha$ is a coefficient that controls the probability assigned to unseen words, and $C$ is the entire corpus. Due to space restrictions, we do not discuss the implementation details of LM-Dirichlet or LM-JM; however, they can be found in \cite{zhai2001study}.

\subsubsection{Query Quality Features}\label{sec:qq}
\label{QQ_features}
We also use a set of metrics that were applied in previous work as estimators for the quality of software artifacts when used as queries for IR \cite{haiduc2012automatic, haiduc2013ICSE}, and were subsequently applied specifically to the task of identifying hard-to-trace artifacts in TLR \cite{mills2017predicting}. We use these metrics as they complement our IR-based features and can give the classifier more contextual information about the link between two artifacts. For example, if the IR rank of an artifact in a potential link is low (i.e., a poor textual match with the other artifact in the link), query quality (QQ) metrics can give an indication of whether this is due to the artifact being generally hard-to-trace, or to the fact that the two artifacts are indeed not related (i.e., linked). 

We adopted all 28 QQ metrics used in previous work in software engineering \cite{mills2017predicting}. The metrics can be split into two main categories: pre-retrieval (21 metrics), which can be applied without running the query and capture general properties of the text found in the artifact, and post-retrieval (7 metrics), which also take into account the ranked list of results returned when running the artifact as a query. Each main category further contains sub-categories, which focus on different properties of a document. Due to space constraints, we refer the interested reader to a previous study that used these metrics for concept location and TLR \cite{mills2017predicting}. 

For each possible link, we applied each pre-retrieval metric to the two documents in the potential link (since each one can be used as a query), resulting in \textit{42 different pre-retrieval QQ features for each link}. Further, we computed each of the seven post-retrieval metrics using five different IR approaches and two different retrieval directions (similarly to the IR-based features described above, post-retrieval QQ metrics also depend on the retrieval direction), resulting in \textit{70 different post-retrieval QQ features for each link}. Note that we do not compute post-retrieval features for LSA or LDA as most of these metrics require some form of document perturbation, which results in the need to re-index the space in which documents are represented for each query. This is particularly cost intensive for these two approaches. Further, document perturbations are performed many times per metric to minimize the effect of non-determinism, which makes it computationally infeasible to use these metrics as features for our representation. 

\subsubsection{Document Statistics Features}
\label{doc_features}
In addition to the aforementioned attributes, we also include some simple document-level features. These metrics are intended to gauge document relevance through term overlap and provide information on the size of documents as a proxy for the information contained in the document. The three document features we use are: the number of unique terms in a document, the total number of terms in a document (including duplicates), and the percentage of overlapping terms between the two documents in a candidate link. Therefore, we compute two features for each artifact in a possible link as well as one feature for the link itself, resulting in a total of \textit{5 features for each link}.

In summary, we extracted 131 features from each potential traceability link: 14 IR-based features, 42 pre-retrieval QQ metrics, 70 post-retrieval QQ metrics, and 5 document features. We normalized each of these to the interval [0,1].

\subsection{Feature Selection}
\label{subsec:selection}
Each feature included in a predictive model's internal representation contains some possibility for error. One mechanism to limit the impact of this potential error is to construct a so-called \textit{parsimonious}  model: one that explains a phenomena with the lowest dimensional internal representation possible. Parsimonious models also limit the effects of spurious correlations, reduce the probability of overfitting, and minimize computation time required to generate the representation. There are many techniques for lowering the dimensionality of a feature space. In this work we consider five algorithms implemented in Weka. The first is Correlation-based Feature Subset Selection (CFS-Subset), which considers the predictive power of each independent variable as well as their mutual correlation. The remaining four algorithms consider the relationship between each independent variable and the dependent (i.e., outcome) variable in terms of: Pearson's Correlation, Gain Ratio, Information Gain, and Symmetrical Uncertainty.

\subsection{Class Imbalance and Data Rebalancing}
\label{subsec:imbalance}
In any given software project, it is expected that the number of possible traceability links is much larger than the number of valid links that actually exist between related artifacts. As such, the boolean classification performed by \approach{} has an inherent \textit{class imbalance}, which can make it difficult to differentiate minority class instances due to insufficient data for learning a pattern \cite{He2009}. Indeed, if there are only two valid links in a total set of 100 possible links, a classifier can optimize accuracy by predicting all links as invalid, unfortunately miss-classifying the two links that are of the highest importance. In the context of our classification, this translates to low recall, which severely impacts the applicability of an automated approach to traceability maintenance. 

To address this problem, rebalancing techniques can be applied to data used to train the classifier, which provides a more equivalent statistical representation of the majority and minority classes. Undersampling and oversampling are two widely used approaches to deal with class imbalance \cite{He2009}. Undersampling involves reducing the majority class by selecting only a subset of its datapoints for training. Oversampling tries to increase the number of datapoints in the minority class, often by artificially creating new datapoints based on those in the original data. We employ both types of sampling methods in our evaluation in order to determine if they help improve the results of \approach{}. Specifically, we use the Synthetic Minority Oversampling TEchnique (SMOTE) \cite{chawla2002smote} and Random Majority Undersampling \cite{He2009}. SMOTE artificially constructs minority datapoints by interpolating data from known minority cases. In Random Majority Undersampling, majority class instances are randomly selected until a sufficiently large sample is obtained. In addition to these two techniques, we also consider a mixed approach, where we apply half the minority class boost used for full SMOTE, and then apply random majority undersampling to achieve in the end two classes that have approximately equal representation in the training data.

\subsection{Classification Algorithms}
Finally, we also need to select a machine learning classification algorithm that is able to best predict the validity of a potential link. We consider a wide range of algorithms from families previously used for software engineering applications \cite{giger2012method, seiffert2014empirical} in order to empirically determine which algorithm is most suitable. Specifically, in this paper we consider k-Nearest Neighbors with $k=5$ (5NN),  N{\"a}ive Bayes (NB), Logistical Regression, Random Forest (RF), Support Vector Machines (SVM), and a Voting ensemble classifier which combines all of the other algorithms. Due to space constraints we do not discuss the implementation of each of these algorithms. Instead we direct the interested reader to a 2010 review of these algorithms \cite{khan2010review} for a detailed discussion. Note that for our analysis we use the standard Weka implementation of each algorithm without any special parameter tuning\footnote{We used LibSVM as the implementation of the Supported Vector Machine classifier}. This allows us to compare classification algorithms in their default state, realizing that additional tuning can further improve performance.

\section{Study Design}
\label{sec:methodology}
We performed an empirical study with two main \textit{goals}. The first is to empirically determine the \textit{best configuration for \approach{}} in terms of its ability to automatically support traceability maintenance. The second is to \textit{compare the best configuration of \approach{} to popular IR approaches} that have been previously applied to traceability link recovery and maintenance \cite{marcus2005recovery, abadi2008traceability, asuncion2010software, oliveto2010equivalence}. In the context of our study, we specifically aim to answer the following research questions:

\newcommand{\rqone}{What is the best performing configuration for \approach?} 
\newcommand{\rqtwo}{Does \approach{} provide superior support for automated traceability maintenance compared with standard IR approaches?}

\textbf{RQ$_{1}$}: \textbf{\emph{\rqone}} Given our design space, defined in Table \ref{tab:dspace}, we aim to determine the combination of feature selection, rebalancing technique, and classification algorithm that leads to the best performance of \approach{}. We measure performance in terms of F-score, computed as the harmonic mean between precision and recall. We aim to balance recall and precision in order to maximize the number of valid links retrieved by the approach while minimizing the number of false positives contained in the result set.

\textbf{RQ$_{2}$}: \textbf{\emph{\rqtwo}} In RQ$_1$, we establish a baseline, default configuration for \approach{} chosen to have the highest performance across all datasets under consideration in this study. In RQ$_2$, we compare the performance of that default configuration with IR approaches, which represent the most commonly used techniques for traceability link recovery \cite{borg2014recovering}. In order to provide a conservative analysis of the performance of this default configuration, we directly compare it with the IR technique chosen from the seven approaches in \ref{sec:ir-features} to have the highest F-score for each dataset under consideration.

\begin{table*}[tb]
	\centering
	\caption{Design space used for the implementations of \approach{}}
	\label{tab:dspace}
	\begin{tabular}{lll}
		\toprule
		\textbf{Categories}                                                                       & \textbf{Variable} & \textbf{Notes}                                                                                                                                                  \\ \midrule
		\multirow{6}{*}{\begin{tabular}[c]{@{}l@{}}Classification \\ algorithms\end{tabular}}         & 5NN               & K-nearest neighbors classifier using K=5                                                                                                                        \\
		& Logistic          & Multinomial logistic regression model with a ridge estimator                                                                                                    \\
		& NaïveBayes        & Naïve Bayes classifier using estimator classes                                                                                                                 \\
		& Random Forest     & Classifier that uses a multitude of random forests                                                                                                              \\
		& SVM               & Supported Vector Machine classifier                                                                                                                             \\
		& Voted             & Ensembled classifier that averages the output of the previous 5 classifiers to classify links                                                                   \\ \midrule
		\multirow{6}{*}{\begin{tabular}[c]{@{}l@{}}Feature\\ Selection\\ techniques\end{tabular}} & none              & All attributes are used (i.e., no feature selection is performed)                                                                                               \\
		& cfs               & \begin{tabular}[c]{@{}l@{}}Evaluates the worth of a subset of attributes by considering the individual predictive ability of each feature\\ along with the degree of redundancy between them\end{tabular} \\
		& correlation       & Evaluates the worth of an attribute by measuring the correlation (Pearson's) between it and the class                                                           \\
		& gainRatio         & Evaluates the worth of an attribute by measuring the gain ratio with respect to the class                                                                       \\
		& infoGain          & Evaluates the worth of an attribute by measuring the information gain with respect to the class                                                                 \\
		& symmetrical       & Evaluates the worth of an attribute by measuring the symmetrical uncertainty with respect to the class                                                          \\ \midrule
		\multirow{4}{*}{\begin{tabular}[c]{@{}l@{}}Rebalancing\\ techniques\end{tabular}}       & none              & No rebalancing technique is applied                                                                                                                             \\
		& undersampling     & Random undersampling is applied to the majority class until it is as small as the minority class                                                                                                          \\
		& smote         & SMOTE is applied to the minority class until it is as large as the majority class                                                                                                                        \\
		& 5050              & \begin{tabular}[c]{@{}l@{}} First, SMOTE is applied half-way on the minority class, then random undersampling is applied to the majority class\\ until the two classes are equal in size\end{tabular} \\
		 \bottomrule
	\end{tabular}
\vspace{-.6cm}
\end{table*}

\subsection{Data Collection}
To answer both research questions, we use a diverse collection of datasets available from the Center of Excellence for Software and Systems Traceability (CoEST) \cite{CoEST} that have been previously used to evaluate new ideas and techniques in the area of traceability \cite{hayes2003improving, DBLP:journals/tse/HayesDS06, cleland2010machine, ali2011requirements, gethers2011integrating}. In total, we use 11 datasets that involve eight different types of artifacts, and are extracted from six different software projects. In total, the datasets have 32,616 possible links between pairs of artifacts, of which 2,600 (7.97\%) are actually valid and 30,016 are invalid (92.03\%). Since the artifacts in these datasets are all either natural language-based (in English or Italian) or source code, we applied the typical preprocessing steps employed in traceability link recovery on this type of artifacts \cite{borg2014recovering}. First, all artifacts were preprocessed to split identifiers on camelCase and underscores ("\_"), then we removed common English (or Italian) words and Java keywords, and finally we stemmed the remaining words to their root form. 

Table \ref{tb:datasets} depicts the datasets used in our evaluation, containing the number of invalid traceability links, the number of valid traceability links, and the artifact types present in each dataset. Note that, as discussed in section \ref{subsec:imbalance}, the data is highly imbalanced, with an average invalid-to-valid link ratio of approximately 12 : 1.

\vspace{-.6cm}
\begin{center}
\begin{threeparttable}[tb]
	\small
	\caption{Datasets used in the evaluation}
	\label{tb:datasets}
	\begin{tabular}{lrrl}
		\toprule	
		\textbf{System} 	&\textbf{\pbox{10cm}{\RaggedLeft Invalid \\ Links}} & \textbf{Valid Links} & \textbf{Artifact Types\tnote{\textdagger}} \\			
		\midrule
		eAnci				& 	7091		& 554 (7.25\%) 				&   UC, CC			\\		
		EasyClinic			&	1317		&	93	(6.60\%)			&		UC, CC			\\
		EasyClinic			&	871			&	69	(7.34\%)			&		ID, CC			\\
		EasyClinic			&	1177		& 83	(6.59\%)			&		ID, TC			\\			
		EasyClinic			&	574			&	26	(4.33\%)			&		ID, UC			\\
		EasyClinic 			&	2757		& 204 (6.89\%)				& 	TC, CC			\\
		EasyClinic			&	1827		&	63	(3.33\%)			&		TC, UC			\\
		eTour				&	6363		& 365 (5.43\%) 				&  	UC, CC			\\
		iTrust				&	1493		&	58 (3.74\%)				&		UC, CC			\\
		MODIS				&   890			&	41 (4.40\%)				&		HighR, LowR			\\
		SMOS				&	5656		& 1044 (15.58\%)			&		UC, CC			\\
		\midrule
		\textbf{Total}	&	\textbf{30016}	&	\textbf{2600 (7.97\%)}			&						 \\
		\bottomrule
	\end{tabular}
	\begin{tablenotes}
		\item[\textdagger]HighR = High-level Requirements, LowR = Low-level Requirements, UC = Use Cases, CC = Code Classes, ID = Interaction Diagrams, TC = Test Cases.
	\end{tablenotes}			
\end{threeparttable}
\end{center}

\subsection{Answering \textbf{RQ$_{1}$} - Determining the best configuration of \approach{}}
In the first set of experiments, we exhaustively searched the design space for a combination of feature selection, rebalancing technique, and classification algorithm that provides the best performance in terms of F-score across all datasets in the study. We consider this configuration as a baseline \approach{} implementation for RQ$_2$. To perform the exhaustive search, we implemented \approach{} using each possible configuration in the design space. We then evaluated each implementation of \approach{} with every dataset independently, running 10-fold cross validations 50 different times and averaging the results. We perform 10-fold cross validations 50 times to account for randomization introduced by the rebalancing techniques and the stratified sampling used in cross validation.

Finally, we collected confusion matrices for each implementation, establishing which configuration achieved the highest average F-score across datasets. We use F-score because it provides a balance between recall (i.e., the ability to extract valid links) and precision (i.e., the ability to minimize false positives). After establishing the best configuration for \approach, we used the Mann-Whitney U test (with Holm-Bonferroni correction) to determine whether there is a statistically significant difference between that configuration and the others. We also established effect size using Cliff's delta.

\subsection{Answering \textbf{RQ$_{2}$} - Comparing \approach{} to IR} \label{sec:rq2-method}
In the second set of experiments, we determine the impact of our contribution by comparing the baseline implementation of \approach{} (determined in RQ$_{1}$) with IR approaches previously used to support TLR. Specifically, we compare against the IR technique among the seven described in Section \ref{sec:qq} that achieves the highest F-score for each dataset. This leads to a conservative comparison of our baseline configuration.

One important aspect to take into consideration for this comparison is the different nature of the output offered by \approach{} and an IR technique: while the former provides a static set of traceability links predicted as valid, the latter offers a \textit{ranked} list of potential traceability links. To directly compare these two different approaches, we considered each IR technique as a classifier, in which all traceability links in the top $k$ positions are classified as \emph{valid links}, and those below position $k$ are classified as \emph{invalid links}. Henceforth, we will refer to the value of $k$ as the \textit{cut-point}. Note that this interpretation is specifically for experimental purposes and has previously been used to evaluate similarity-based classifiers in TLR \cite{cleland2010machine}. Since \approach{} predicts a certain number $N$ of valid traceability links per dataset, we compared \approach{} with each IR technique for every dataset using $N$ as the cut-point.

Each IR technique provides a single value for each performance metric; for \approach{}, we run 50 trials of 10-fold cross validation on each dataset in order to mitigate the effect of sampling bias and then average the results. We determined if the resulting distributions of precision, recall, and F-score obtained by \approach{} per dataset were normally distributed using the Shapiro-Wilk test with a significance level of 0.01. Next, to compute significance between the distributions provided by \approach{} and the single value of each metric provided by the best IR technique for each dataset, we used either the one-sample t-test (if the observations were normally distributed) or the one-sample Mann-Whitney U test (if the observations were non-normally distributed). Finally, we adjusted the obtained p-values for each metric using the Holm-Bonferroni correction for each dataset. In either case we compute effect sizes by normalizing the difference between the distribution mean and the IR metric by the distribution's standard deviation. This calculation penalizes non-normally distributed samples, rewarding those with high mean and low standard deviation.

The results of the experiments and the data used in this study are publicly available in a replication package \cite{replication}.


\section{Results}
\label{sec:results}

\subsection{\textbf{RQ$_1$} - Determining the best configuration of \approach{}}

\begin{table*}[tb]
	\centering
	\begin{threeparttable}
		\caption{Average F-score (in percentage) achieved by the implementations of \approach{} across all datasets. The best configuration is in underlined, bold font.}
		\label{tab:rq1}
		\begin{tabular}{llllllll}
			\toprule
			\multirow{2}{*}{\textbf{\begin{tabular}[c]{@{}l@{}}Rebalancing\\ Technique\end{tabular}}} & \multirow{2}{*}{\textbf{\begin{tabular}[c]{@{}l@{}}Feature\\ Selection\end{tabular}}} & \multicolumn{6}{c}{\textbf{Classifier}}                                                                       \\ \cmidrule(l){3-8}
			&                                                                                       & \textbf{5NN} & \textbf{Logistic} & \textbf{NaiveBayes} & \textbf{RandomForest} & \textbf{SVM} & \textbf{Vote} \\
			\hline
			\multirow{6}{*}{\textbf{none}}                                                            & \textbf{none}                                                                         & 47.43$^*$    & 50.19$^*$         & 39.49$^*$           & 67.18$^\dagger$       & 0.00$^*$        & 55.96$^*$     \\
			& \textbf{cfs}                                                                          & 59.72$^*$    & 40.25$^*$         & 39.22$^*$           & 63.14$^*$             & 0.79$^*$     & 53.84$^*$     \\
			& \textbf{correlation}                                                                  & 47.43$^*$    & 50.19$^*$         & 39.49$^*$           & 67.22$^\dagger$       & 0.00$^*$        & 56.06$^*$     \\
			& \textbf{gainRatio}                                                                    & 61.22$^*$    & 61.00$^*$            & 40.17$^*$           & 72.03                 & 0.00$^*$        & 66.84$^*$     \\
			& \textbf{infoGain}                                                                     & 61.22$^*$    & 61.00$^*$            & 40.17$^*$           & 72.29                 & 0.00$^*$        & 66.96$^*$     \\
			& \textbf{symmetrical}                                                                  & 61.22$^*$    & 61.00$^*$            & 40.17$^*$           & 72.07                 & 0.00$^*$        & 66.89$^*$     \\	\midrule
			\multirow{6}{*}{\textbf{undersampling}}                                                   & \textbf{none}                                                                         & 31.18$^*$    & 34.60$^*$          & 36.13$^*$           & 51.37$^*$             & 31.22$^*$    & 38.24$^*$     \\
			& \textbf{cfs}                                                                          & 39.88$^*$    & 37.65$^*$         & 35.65$^*$           & 47.43$^*$             & 34.77$^*$    & 38.35$^*$     \\
			& \textbf{correlation}                                                                  & 31.18$^*$    & 34.59$^*$         & 36.13$^*$           & 51.42$^*$             & 31.41$^*$    & 38.28$^*$     \\
			& \textbf{gainRatio}                                                                    & 37.63$^*$    & 38.05$^*$         & 37.81$^*$           & 51.38$^*$             & 35.69$^*$    & 41.82$^*$     \\
			& \textbf{infoGain}                                                                     & 37.63$^*$    & 38.05$^*$         & 37.81$^*$           & 51.34$^*$             & 35.69$^*$    & 41.83$^*$     \\
			& \textbf{symmetrical}                                                                  & 37.63$^*$    & 38.05$^*$         & 37.81$^*$           & 51.41$^*$             & 35.69$^*$    & 41.83$^*$     \\	\midrule
			\multirow{6}{*}{\textbf{smote}}                                                           & \textbf{none}                                                                         & 56.19$^*$    & 56.31$^*$         & 38.05$^*$           & 74.80                  & 46.77$^*$    & 56.94$^*$     \\
			& \textbf{cfs}                                                                          & 54.07$^*$    & 41.04$^*$         & 37.46$^*$           & 62.74$^*$             & 41.44$^*$    & 45.42$^*$     \\
			& \textbf{correlation}                                                                  & 56.15$^*$    & 56.33$^*$         & 38.05$^*$           & \textbf{\underline{75.18}}  & 46.99$^*$    & 56.99$^*$     \\
			& \textbf{gainRatio}                                                                    & 63.10$^*$     & 58.05$^*$         & 39.74$^*$           & 73.89                 & 47.68$^*$    & 57.74$^*$     \\
			& \textbf{infoGain}                                                                     & 63.10$^*$     & 57.95$^*$         & 39.75$^*$           & 73.88                 & 47.67$^*$    & 57.74$^*$     \\
			& \textbf{symmetrical}                                                                  & 63.09$^*$    & 58.06$^*$         & 39.77$^*$           & 73.85                 & 47.69$^*$    & 57.77$^*$     \\	\midrule
			\multirow{6}{*}{\textbf{5050}}                                                            & \textbf{none}                                                                         & 49.59$^*$    & 51.19$^*$         & 38.03$^*$           & 72.09$^\dagger$       & 43.70$^*$     & 53.36$^*$     \\
			& \textbf{cfs}                                                                          & 51.24$^*$    & 40.80$^*$          & 37.35$^*$           & 61.47$^*$             & 40.34$^*$    & 44.33$^*$     \\
			& \textbf{correlation}                                                                  & 49.60$^*$     & 51.17$^*$         & 38.02$^*$           & 72.33$^\dagger$       & 43.97$^*$    & 53.38$^*$     \\
			& \textbf{gainRatio}                                                                    & 57.04$^*$    & 56.64$^*$         & 39.67$^*$           & 70.62$^\dagger$       & 45.55$^*$    & 55.01$^*$     \\
			& \textbf{infoGain}                                                                     & 57.10$^*$     & 56.63$^*$         & 39.69$^*$           & 70.64$^\dagger$       & 45.58$^*$    & 54.99$^*$     \\
			& \textbf{symmetrical}                                                                  & 57.04$^*$    & 56.64$^*$         & 39.68$^*$           & 70.60$^\dagger$        & 45.55$^*$    & 55.02$^*$    \\ \bottomrule
		\end{tabular}
		\begin{tablenotes}
			\item * = implementations performing statistically significantly worse than the best configuration (0.01 significance level), with a medium or large effect size;
			`$\dagger$' = implementations performing statistically significantly worse than the best configuration (0.01 significance level), with a small effect size.
		\end{tablenotes}
	\end{threeparttable}
	\vspace{-.3cm}
\end{table*}

Table \ref{tab:rq1} shows the average F-score achieved by each combination of parameters in our design space across all of the datasets in our study. The configuration with the highest F-score is displayed in bold with an underline, and is the implementation chosen as a default configuration for use in RQ$_2$. The configurations with statistically significantly lower F-scores (at the .01 significance level) compared to the best performing configuration are marked with a superscript symbol, where an asterisk (`*') indicates a medium or large effect size and a cross (`$\dagger$') indicates a small effect size. The results indicate that \textit{Random Forrest (RF)} outperforms all other classification algorithms across all other dimensions, and achieves the best results when using the \textit{Pearson correlation }for feature selection and \textit{SMOTE} for data rebalancing.

The findings also show that the choice of \textit{balancing technique} is important, and that undersampling provides the lowest F-scores for all classification algorithms but SVM, across all feature selection techniques. When trying to learn more about this drop in F-score, we noticed that for all configurations not involving SVM, random undersampling boosted recall, but also drastically lowered the precision. Therefore, while undersampling allows \approach{} to retrieve more valid links in those cases, they are returned in addition to many false positives. One reason for this could be that, by reducing the number and diversity of false links to learn from, there were not enough false link instances for \approach{} to learn all the nuances found in the large number of false links in the dataset, therefore misclassifying many of them as true links.

The results also show that completely rebalancing the classes with SMOTE leads to similar F-scores to configurations in which no rebalancing is performed. When analyzing the recall and precision in more detail, we found that SMOTE increases \approach's recall substantially, from 57\% (with no rebalancing) to 76\% in the case of our selected configuration, while maintaining a precision of 75\%. On the other hand, the selected configuration without any balancing results in a higher precision of 86\%, but at 57\% recall only slightly more than half of the valid links are extracted. While there are scenarios in which recall should be prioritized over precision or vice versa, in this study we are interested in optimizing the two together. Therefore, practitioners interested in implementing \approach{} should choose an implementation that best suits their needs. For example, in some applications an approach with high recall can be used to reduce the number of document pairs to inspect, while ensuring that a minimal number of valid links are missed. Other applications might call for an approach that maximizes confidence in \approach's predictions by maximizing precision at the cost of overlooking some valid links.

When comparing \textit{feature selection techniques} for RF with SMOTE, which achieves the best results overall, there are small differences in F-score between the parameter choices except for the CFS-Subset algorithm, which achieves a significantly lower F-score. Selection based on Pearson's correlation provides the best F-score by a narrow margin, and has higher precision than the other options at a slightly reduced recall. There are no statistical differences between configurations using SMOTE with the RF algorithm and any feature selection other than CFS-Subset. Therefore, from a practical point of view, one could also choose the implementation that requires the smallest number of features but statistically provides the same performance. Table \ref{tab:selected-features} shows the number of features selected by each of the feature selection techniques for each dataset. As the table indicates, the number of features selected is dependent on the system. For this study, however, our goal was to find the best single configuration across all systems. As such, we strictly focus on the implementation that provides the largest average F-score across all systems, i.e., the Pearson's correlation-based feature selection with full SMOTE and the RF algorithm.
 
In summary, these findings suggest that the most important consideration in designing a \approach{} implementation is the classification algorithm, where RF shows statistically significantly higher F-scores compared to other algorithms across all other dimensions. Secondarily, rebalancing and feature selection are considerations that can be used to arrive at an implementation that favors either recall or precision based on contextual need with the smallest number of features for a given project. 

\vspace{.4cm}
\noindent\fbox{%
	\parbox{\dimexpr0.97\columnwidth\relax}{%
	 \textit{Correlation-based feature selection} with \textit{full SMOTE rebalancing} and the \textit{Random Forest classification algorithm} are the best \approach{} configuration for our dataset based on F-score.
	}%
}
\vspace{-0.1cm}

\begin{table}[tb]
	\centering
	\caption{Number of features selected by each technique per dataset. The initial number of features is 131}
	\label{tab:selected-features}
	\begin{tabular}{lrrrrr}
		\toprule
		\textbf{Dataset}           & \textbf{cfs} & \textbf{corr} & \textbf{gainR} & \textbf{infoG} & \textbf{symm} \\ \midrule
		eAnci(CC-UC)      & 8   & 127         & 125       & 125      & 125         \\
		EasyClinic(CC-UC) & 10  & 127         & 58        & 58       & 58          \\
		EasyClinic(ID-CC) & 6   & 127         & 40        & 40       & 40          \\
		EasyClinic(ID-TC) & 9   & 123         & 102       & 102      & 102         \\
		EasyClinic(ID-UC) & 9   & 125         & 19        & 19       & 19          \\
		EasyClinic(TC-CC) & 10  & 126         & 70        & 70       & 70          \\
		EasyClinic(TC-UC) & 6   & 125         & 69        & 69       & 69          \\
		eTour(CC-UC)      & 10  & 127         & 72        & 72       & 72          \\
		iTrust(CC-UC)     & 9   & 126         & 72        & 13       & 13          \\
		Modis(HighR-LowR)   & 3   & 127         & 65        & 65       & 65          \\
		SMOS(CC-UC)       & 11  & 125         & 76        & 76       & 76          \\ \bottomrule
	\end{tabular}
\begin{tablenotes}
	\item HighR = High-level Requirements, LowR = Low-level Requirements,\\ UC = Use Cases, CC = Code Classes, ID = Interaction Diagrams,\\TC = Test Cases.
\end{tablenotes}
\vspace{-.5cm}
\end{table}

\subsection{\textbf{RQ$_2$} - Comparing \approach{} to IR}

Table \ref{tab:r2} shows a comparison of precision, recall, and F-score between our derived baseline \approach{} implementation and the best performing IR technique for each dataset from the set of seven techniques considered in our study. Again, the best performing IR approach was chosen by selecting the one with the highest F-score, as done for the default configuration of \approach{}. For IR, the table shows precision, recall and F-score at K, where K is the cut-point chosen to be the number of links \approach{} predicts to be valid (which allows for a direct comparison between the approaches, as described in section III-C). The IR techniques that performed statistically significantly worse than \approach{} with a significance level of 0.01 and large effect sizes or greater are denoted with an asterisk (`*'). Interestingly, VSM, one of the most basic IR techniques considered in this study, had the highest F-score in nine of twelve systems, whereas LDA and LM-JM did not have the highest F-score for any of the cases. This is consistent with previous work on IR-based TLR \cite{oliveto2010equivalence}.

Our baseline \approach{} implementation outperforms even the best IR approach in terms of precision, recall, and F-score for each of the 11 datasets in this study. However, we note that the performance of both \approach{} and IR is dependent on the dataset. For six datasets, we achieve higher than 70\% recall and precision, while the lowest recall and precision for \approach{} is 56\% and 59\% respectively for EasyClinic(ID-UC). For IR, recall and precision higher than 60\% is only achieved in one case (EasyClinic(CC-UC)), and the lowest recall and precision are 30\% and 28\% respectively, more than 25 percentage points lower than that of \approach{}.

There is some variability in the performance improvement provided by \approach{}. For example, in the case of EasyClinic(CC-UC) and iTrust, \approach{} outperforms VSM by less than five percentage points. However, on the other end of the spectrum, \approach{} is able to extract more than 90\% of the valid links (with a maximum of 99\%) for three of the EasyClinic datasets with higher than 90\% precision. This is a substantial improvement over IR approaches, which never achieve more than 60\% recall or precision for those datasets. Similarly, \approach{} improves performance for eAnci in terms of each metric by more than 40 percentage points. Further, in six of the datasets, \approach{} outperforms IR by more than 30 percentage points for all three metrics. Overall, our baseline \approach{} configuration outperforms IR by more than 26 percentage points in average precision, recall, and F-score. Finally, the performance of \approach{} is statistically significantly better than IR at the .01 significance level for all three metrics for each dataset, with a large effect size or greater in each case. Therefore, while many of the features used by our baseline implementation of \approach{} are derived directly from IR techniques, our findings indicate that the power of \approach{} may come from combining this type of information with the other features used. 

In summary, the results of our study indicate that even our baseline implementation of \approach{} outperforms traditional IR approaches for all 11 datasets in terms of all three performance metrics. Further, while the improvement in performance is limited in some cases, in the majority of cases \approach{} improves each measure by more than 10 percentage points. Moreover, these findings are statistically significant at the .01 significance level, with a large effect size or greater. Considering all datasets in the study, \approach{} outperforms even the best IR for each dataset by more than 26 percentage points in terms of average precision, recall, and F-score. Given that we did not optimize \approach{} per dataset, but rather used a baseline, general implementation across all systems, we anticipate that further refining our parameters per dataset will lead to even better results for \approach{} in future work. 

\vspace{.4cm}
\noindent\fbox{%
	\parbox{\dimexpr0.97\columnwidth\relax}{%
	\approach{} significantly outperforms the best Information Retrieval approaches in terms of precision, recall, and F-score in all datasets.}%
}
\vspace{0.0cm}

\begin{table*}[tb]
	\centering
	\begin{threeparttable}[b]
\caption{Precision, recall and F-score (in percentage) for \approach{} compared to the IR technique with the highest F-score for each dataset.}
\label{tab:r2}
\begin{tabular}{lccccccc}
	\toprule	
	\multicolumn{1}{c}{\multirow{2}{*}{\textbf{Dataset}}} & \multicolumn{3}{c}{\textbf{TRAIL}}                                                                                  & \multicolumn{4}{c}{\textbf{Information Retrieval}}                                                                                                               \\	\cmidrule(lr){2-4} \cmidrule(l){5-8}
	\multicolumn{1}{c}{}                                  & \multicolumn{1}{c}{\textbf{Precision(\%)}} & \multicolumn{1}{c}{\textbf{Recall(\%)}} & \multicolumn{1}{c}{\textbf{F-score(\%)}} & \multicolumn{1}{c}{\textbf{Best IR}} & \multicolumn{1}{c}{\textbf{Precision@K(\%)}} & \multicolumn{1}{c}{\textbf{Recall@K(\%)}} & \multicolumn{1}{c}{\textbf{F-score@K(\%)}} \\ \midrule
	eAnci(CC-UC)             & 75.46          & 80.53          & 77.91          & VSM       & $28.21^*$          & $30.14^*$          & $29.14^*$          \\
	EasyClinic(CC-UC)        & 64.06          & 71.46          & 67.54          & VSM       & $61.54^*$          & $68.82^*$          & $64.97^*$          \\
	EasyClinic(ID-CC)        & 72.56          & 73.19          & 72.83          & VSM       & $58.57^*$          & $59.42^*$          & $58.99^*$          \\
	EasyClinic(ID-TC)        & 94.14          & 92.75          & 93.43          & LSA          & $58.54^*$          & $57.83^*$          & $58.18^*$          \\
	EasyClinic(ID-UC)        & 59.27          & 56.23          & 57.59          & VSM       & $48.00^*$             & $46.15^*$          & $47.06^*$          \\
	EasyClinic(TC-CC)        & 95.96          & 99.03          & 97.47          & VSM       & $44.08^*$          & $45.59^*$          & $44.82^*$          \\
	EasyClinic(TC-UC)        & 93.39          & 95.62          & 94.47          & LM-Dirichlet & $55.38^*$          & $57.14^*$          & $56.25^*$          \\
	eTour(CC-UC)             & 57.17          & 64.98          & 60.82          & VSM       & $49.40^*$           & $56.16^*$          & $52.56^*$          \\
	iTrust(CC-UC)            & 56.79          & 65.76          & 60.92          & VSM       & $52.94^*$          & $62.07^*$          & $57.14^*$          \\
	Modis(HighR-LowR)          & 65.88          & 62.93          & 64.32          & VSM       & $32.50^*$           & $31.71^*$          & $32.10^*$           \\
	SMOS(CC-UC)              & 87.07          & 73.53          & 79.73          & BM25         & $37.76^*$          & $31.90^*$           & $34.58^*$          \\ \midrule
	\textbf{Average}         & \textbf{74.70} & \textbf{76.00} & \textbf{75.18} & \textbf{}    & \textbf{47.90} & \textbf{49.72} & \textbf{48.71} \\ \bottomrule
	
\end{tabular}
\begin{tablenotes}
	\item * = statistical significance (0.01 significance level) with a large effect size or greater.
\end{tablenotes}
	\end{threeparttable}
\vspace{-.5cm}
\end{table*}

\section{Threats to Validity}
\label{sec:threats}
\textit{Construct validity} refers to how well the metrics used for evaluation truly capture what they were intended to. We mitigated this risk in several different ways. First, we measured performance of both \approach{} and IR techniques throughout the analysis using metrics that have been commonly used in software engineering research, and specifically in TLR studies \cite{Cleland-Huang2007,borg2014recovering}. Second, for both RQ$_1$ and RQ$_2$, we considered the effect of randomization on the performance metrics as it is possible using only one trial to obtain good results purely by chance. Therefore, we performed 50 trials of \approach{} for each experiment, averaging to aggregate our final results. Finally, we employ statistical tests where applicable, adding rigor to the analysis and empirically supporting our claims. 

\textit{Internal validity} refers to how well a study mitigates multiple independent (i.e., conflating) variables from interfering with inferences on the dependent variable. We mitigate these risks in RQ$_1$ by exhaustively searching the design space, systematically varying independent variables one at a time. Further, in answering this research question we derive a baseline implementation of \approach{} which is then compared to the best performing IR approach for each dataset. Finally, in RQ$_2$, we carefully constructed a methodology that allows direct comparisons between \approach{} and IR despite their fundamentally different approaches to extracting traceability links and corresponding output.

\textit{External validity} investigates how well the findings of a study can be generalized. For this study we mitigated this risk by considering a diverse set of 11 datasets extracted from six different software systems and eight types of artifacts and unique artifact combinations, involving more than 30,000 potential traceability links. Moreover, these are all datasets curated by the research community, which have been used in previous traceability studies.

\section{Related Work}
\label{sec:related}
\label{section:related}
Broadly speaking, software traceability covers a wide range of contexts linking abstract concepts such as architectural tactics and non-functional requirements (NFRs) \cite{cleland2007automated, mirakhorli2012tactic} or software artifacts like use cases and source code \cite{antoniol2000tracing, binkley2014learning}. Significant previous work has been done to increase the level of automation available for establishing and maintaining these links, most commonly using IR techniques to rank artifacts or candidate links based on document similarity. Here we discuss four main categories of approaches: IR, machine learning or classification, event-driven, and model-based approaches.

Applications of IR to TLR are the most related to this work, and began with probabilistic \cite{antoniol2000tracing, antoniol1999recovering} and vector space models \cite{antoniol2002recovering} to retrieve links between source code and documentation, as well as source code and requirements. Additionally, other IR approaches such as Latent Semantic Analysis (LSA) \cite{marcus2005recovery}, probabilistic LSA (pLSA) \cite{abadi2008traceability}, Jensen-Shannon (JS)\cite{abadi2008traceability}, Latent Dirichlet Allocation (LDA) \cite{asuncion2010software}, and proximity-based VSM \cite{kong2011proximity} have also been applied directly to the TLR task. 
Oliveto et al. performed an empirical study of IR methods for TLR \cite{oliveto2010equivalence} comparing VSM, LSA, LDA, and JS via Principal Component Analysis, showing that VSM, LSA, and JS capture similar information, while information captured by LDA is unique. 
Learning to Rank has been applied to improve IR approaches \cite{binkley2014learning}, deriving a more accurate ranking of document similarities by combining information from various IR approaches into a single list. More recently, Falessi et al. \cite{falessi2016estimating} used machine learning classifiers to estimate the number of valid links remaining in a set of candidate links returned by IR techniques.

There are also a large number of techniques for increasing the performance of IR approaches to TLR by augmenting document information with key-phrases and technical terms through a thesaurus \cite{hayes2003improving}, applying alternative term weighting strategies \cite{settimi2004supporting, sundaram2010assessing}, and smoothing filters \cite{de2011improving}. Additionally, Cleland-Huang et al. \cite{cleland2005utilizing} propose three different improvement strategies based on artifact hierarchies, clustering, and graph pruning. Further, previous research has considered the selection of specific IR-based infrastructures for TLR through the use of genetic algorithms \cite{lohar2013improving}. For a more in-depth overview of related work on IR in TLR, we direct the reader to a systematic mapping study of the field \cite{borg2014recovering}. Our work differs from this previous work as we leverage IR techniques to generate features for our classifier rather than applying them directly to the task of establishing traceability links. 

There is also existing work specifically focused on classification and traceability, but significantly different than our approach. In the area of requirements engineering, Cleland-Huang et al. proposed a probabilistic classifier trained on a set of indicator words for NFRs \cite{cleland2007automated}, which was shown to outperform traditional machine learning classifiers \cite{jalaji2006making}. This classifier was applied to traceability specifically for linking regulatory codes to project-specific requirements \cite{cleland2010machine} and architectural tactics to source code \cite{mirakhorli2012tactic}. A semi-supervised, iterative approach utilizing Expectation Maximization has also been proposed \cite{casamayor2010identification}. Further, Asuncion et al. constructed an automatic approach called TRASE using probabilistic topic modeling with a modified form of LDA to automatically infer traceability links \cite{asuncion2010software}.

Our work differs from these classification schemes in a couple of ways. The approach by Cleland-Huang et al. \cite{cleland2010machine} requires the identification of a set of high level concepts that can be represented by an indicator term set. Our approach does not require this intermediate step and is able to directly classify links between artifacts. Furthermore, their approach uses a threshold for classification, similar to IR-based approaches, while our classification criteria are left entirely to the machine learning algorithm, making it fully automatic. Our approach differs from that by Casamayor et al. \cite{casamayor2010identification} in that they provide an iterative approach based on user feedback for requirement classification, while we propose a customizable, yet automatic approach for the classification of traceability links, and in a more generalized context. Compared to TRASE \cite{asuncion2010software}, TRAIL uses LDA as a single IR feature (along many others) to determine document similarity, which is then used by the larger classification model to learn relationships between artifacts.

While the aforementioned approaches can be applied to the maintenance problem by using them to re-establish traceability after changes are made, there have also been event-driven techniques specifically for maintaining existing traceability links. Trigger events for re-evaluating links during system evolution were suggested by Chen and Chou as early as 1999 \cite{chen1999consistency}. Cleland-Huang et al. \cite{cleland2003automating} followed with an approach based on an event server intended to notify stakeholders when related artifacts are updated and traceability information might be stale. Murta et al. \cite{murta2008continuous} use a rule-based approach to ensure that traceability links are consistent throughout the evolution process specifically for maintaining links between architectural concepts and implementing code. M\"{a}der et al. \cite{mader2008rule} provide a modeling plugin that detects high-level operations on software models and propagates appropriate updates to underlying traceability links with related artifacts. Similarly, TraceMAINTAINER \cite{mader2008enabling, mader2009enabling} is a semi-automated maintenance system that also uses events and a rules engine to perform updates. While these approaches specifically seek to address the maintenance problem, they have some significant limitations. First, they typically require a set of rules to be established by which updates are propagated when an event is raised. Second, these approaches often seek to prevent trace decay in which existing links become stale. Therefore, they do not address the situation in which new artifacts are created and must be included in an existing web of traceability links. Our work differs from these approaches in that we use historical training data to maintain links rather than relying on predefined rule sets. Additionally, our approach supports both the situation in which existing links are altered and when new links are established as artifacts are created. 

Finally, there has been a significant amount of work in model-driven engineering around establishing and maintaining traceability links between models. These approaches are related to this work only in that they seek to address the same general problem, but with techniques that are not typically applicable outside of model-driven projects, while \approach{} is fully generalizable. As such, we refer the interested reader to any of several surveys on related topics \cite{galvao2007survey, winkler2010survey, boussaid2017survey}.

\section{Acknowledgements}
This work is supported in part by the National Science Foundation (NSF) grant CCF-1526929.

\section{Conclusion and Future Work}
\label{sec:conclusion}
In this paper, we propose \approach, a novel technique for automating traceability maintenance by considering TLR as a binary classification problem. We address this problem using machine learning algorithms trained on historical traceability information. We empirically derive a high-performance baseline implementation of \approach{} on 11 datasets commonly used to evaluate new approaches to traceability. Moreover, we show that \approach{} significantly outperforms even the best IR approach for each of these datasets as measured by three performance metrics: precision, recall, and F-score.

While a significant improvement over existing IR techniques, \approach{} suffers from some of the same limitations as traditional IR approaches to traceability. For example, this study did not address the so-called \textit{vocabulary mismatch} problem, which arises when different terms are used to express the same ideas between sets of artifacts. 
Recent work in traceability \cite{guo2017tackling} has presented techniques for bridging the term gap between artifact sets. 
Our future work will focus on improving the IR components of our feature set by incorporating state-of-the-art improvements such as this.

Moreover, because at its core \approach{} is a direct application of machine learning classification to the TLR problem, we can apply the most recent advances in this field to expand the framework's capabilities. First, we plan to investigate transfer-learning techniques that make it possible to train a model using historical data from another, similar project. The result is an implementation of \approach{} more widely applicable to greenfield projects with no existing historical data. Further, we will investigate active learning approaches that look to minimize the amount of training data required to generate an effective predictive model for \approach{}. Using these techniques in tandem with transfer learning, we can minimize barriers to adopting \approach{} from a practitioner's perspective. Finally, the existing framework is based on traditional machine learning, which requires feature engineering to derive vector representations of the traceability links. Future work will investigate using deep neural networks capable of automatically extracting vector representations of links for classification, therefore eliminating the need to design features.






\bibliographystyle{IEEEtran}
\bibliography{bibliography}

\end{document}